\documentclass[a4paper,11pt]{article}
\usepackage{comment}
\usepackage{amsfonts,amsmath,amssymb,extarrows,mathtools,graphicx,subfigure,setspace}
\usepackage{cite}
\usepackage{slashed}
\usepackage{color}
\usepackage{jheppub} 
                                       
                     \usepackage{tikz}
                     \usepackage{fancyhdr}
                     \usepackage{xcolor}
                     \usepackage{epstopdf}
                
                     \usepackage{hyperref}
                     \usepackage{cleveref}
                     \usepackage{float}
                     \usepackage{braket}

                     \newcommand{\ud}{\,\mathrm{d}}
                     \newcommand{\la}{\lambda}
                     \newcommand{\deltabar}{{\mkern0.75mu\mathchar '26\mkern -9.75mu\delta}}
                     
                     \usepackage{epsfig,multicol,bbm}
                     
                     \newcommand\fverb{\setbox\pippobox=\hbox\bgroup\verb}
                     \newcommand\fverbit{\egroup\item[\fbox{\unhbox\pippobox}]}
                     
                     \newbox\pippobox

\usepackage[T1]{fontenc} 

\title{\boldmath On the first law of holographic complexity }

\author[]{S. Sedigheh Hashemi,}
\author[]{Ghadir Jafari,}
\author[]{and Ali Naseh}

\affiliation[]{School of Particles and Accelerators, Institute for Research in Fundamental Sciences (IPM),\\ P. O. Box 19395-5531, Tehran, Iran}

\emailAdd{hashemiphys@ipm.ir}
\emailAdd{ghjafari@ipm.ir}
\emailAdd{naseh@ipm.ir}

\abstract{
	In this paper, we examine the proposed first law of holographic complexity through studying different perturbations around various spacetime backgrounds. We present a general expression for the variation of the holographic complexity on arbitrary backgrounds by an explicit covariant computation. Interestingly, the general expression can be written as a function of gravitational conserved charges reminiscent of the first law of thermodynamics.}


\begin{document} 
\maketitle
\flushbottom

\section{Introduction}
In recent years,  {ideas} of quantum information theory have {played an increasingly important role} in {shedding light on dark corners} of quantum field theory (QFT) and quantum gravity. One fascinating concept from quantum information theory that has been recently discussed in the context of QFT is the \emph{quantum circuit complexity}. Loosely speaking, the quantum circuit complexity is defined as the size of the optimal unitary transformation{,} $U_\text{T}${,} which can prepare a target state $\ket{\Psi_{\text{T}}}$ from a reference state  $\ket{\Psi_{\text{R}}}$ by using a set of elementary gates \cite{Aaronson:2016vto,Susskind:2018pmk}.\\

In the context of AdS/CFT correspondence, studies have aimed at understanding the growth of the Einstein-Rosen bridge in the AdS black hole background in terms of quantum complexity in the dual boundary CFT. There are two independent proposals for the gravitational observables, which will be dual to the complexity of a holographic boundary state
\cite{Susskind:2014rva,Brown:2015lvg,Brown:2015bva,Stanford:2014jda,Couch:2016exn}.
One is the complexity=volume ($\mathcal{C}_V$) conjecture \cite{Susskind:2014rva} and the other complexity=action ($\mathcal{C}_A$) conjecture \cite{Brown:2015lvg,Brown:2015bva}. The first proposal{,} states that the complexity of state in the boundary CFT is dual to the volume of the extremal {surface} meeting  the asymptotic boundary on the desired time slice. The second conjecture equates the  complexity of a boundary state with the gravitational action evaluated on the Wheeler-deWitt (WdW) patch,
\begin{equation}\label{1}
\mathcal{C}_{A}(\Sigma) = \frac{I_{\text{WdW}}}{\pi },
\end{equation}
where the WdW patch is defined as the domain of dependence of any bulk Cauchy surface approaching asympotically the time slice $\Sigma$ on the boundary. 
It should be noted that the gravitational observable{s} are sensitive to the bulk physics deep inside a black hole  and exploring their properties is an active area of research 
\cite{Susskind:2015toa,Lehner:2016vdi,Cai:2016xho,Carmi:2016wjl,Abt:2017pmf,Cano:2018aqi,Alishahiha:2015rta,Akhavan:2018wla,Alishahiha:2017hwg,Alishahiha:2018tep}. A{n important} limitation {of this approach} is that the {concept} of circuit complexity {is not yet} well-understood in {the context of interacting QFTs. Therefore, developing the concept of circuit complexity for QFT states, in particular for states of an strongly coupled CFT \cite{Belin:2018bpg,Liu:2019aji}, would be an important task.} Nielsen's geometric approach \cite{Nielsen1133,Nielsen2005AGA} gives a framework to describe the complexity of QFT states.  Based on this approach, one easily finds out that the variation  of circuit complexity only depends on the end point of the optimal trajectory \cite{Bernamonti:2019zyy}, this feature  designated by  Bernamonti et. al \cite{Bernamonti:2019zyy}, as \emph{the first law of complexity}.
The authors examined variation of holographic complexity for two nearby target states where these  states are dual to smooth geometries in the bulk gravitational theory. To be more precise,
they considered  variations of  holographic complexity , $(\delta \mathcal{C}_{A})$, under changing the target state by perturbing {an} AdS background by backreaction of an scalar field.  
The result was that the gravitational contributions to the variation canceled each other, and the final variation came from the scalar field action alone.

{The importance of variation of complexity comparing to complexity itself has different faces. Firstly, the complexity depends on both reference state  $\ket{\Psi_{\text{R}}}$ and the target state $\ket{\Psi_{\text{T}}}$ whereas its variation which will appear in the first law only depend on the  $\ket{\Psi_{\text{T}}}$. The status of the reference state is also not clear in bulk geometry but the target state has good geometrical interpretation, therefore even the holographic calculation of $ \delta\mathcal{C}$ may be more useful than $ \mathcal{C}$. Secondly,
from gravitational point of view the holographic proposals for complexity defined new gauge invariant quantities, so studying their variations under small perturbations may leads to new insights in holography. In fact comparing to first law of entanglement entropy, understanding aspects of first law of complexity may have information about the dynamics of spacetime as (or more than)  first law of entanglement entropy contain information about linearized Einstein's equations. Thirdly, studying this variation can provide a more clear connection between quantum circuit constructions for QFT complexity and its holographic counterpart. Last but not least, the variation of complexity is finite which makes it better defined observable than the complexity itself.}

In the holographic calculations of \cite{Bernamonti:2019zyy}, the background is {taken} to be the AdS space and {the }perturbations were restricted to preserve the spherical symmetry.  In this paper, { we extend the calculations of \cite{Bernamonti:2019zyy} for arbitrary backgound solution, specially the black hole geometry (instead of pure AdS), and in presence of arbitrary perturbations to find the general form of first law of complexity. {This will be an important step toward understanding first law of complexity since the AdS background is dual to the vacuum state in field theory and therefore understanding variation of complexity for thermal states can lead to new insights about this concept. In fact our calculation provide this insight by showing that the conserved charges (defined on the null cones) of space time geometry interestingly appear in the expression for the first law. This observation by itself can help us to learn about spacetime reconstruction where the spacetime can be built up by adding layers of null cones. }

{	
In this paper we study variation  of holographic complexity using the CA proposal. For studying the first law of holographic complexity using the CV proposal see \cite{Bernamonti:2020bcf}. Although the value of complexity in  CA and CV proposals  have some common aspects, its variation  does not seems to have similar behavior in these two proposals. For example, apart the extra logarithmic factors in the UV divergences for the $\mathcal{C}_{A}$,  $\delta \mathcal{C}_{V}$ can becomes negative by adding the relevant operators while $\delta\mathcal{C}_{A}$ is always positive \cite{Bernamonti:2020bcf}.}

This paper is organized as follows:  In section \ref{sec:general}{,} we find {a} general expression for the first law of holographic complexity using the covariant approach for variation of {the} on-shell action evaluated in the WdW patch. In section.\ref{Exapmle} , as an example, we explore the form of first law for the case of some black hole backgrounds. Section \ref{discussion} {includes the} discussion {and} the results. We also devoted two appendices.\ref{AppA},\ref{AppB} in which some details of calculations are provided.
\section{$\delta\mathcal{C}_A$ for general background and arbitrary perturbation}\label{sec:general}
In order to examine  the first law of holographic complexity,
we study the variation of holographic complexity in $\mathcal{C}_A$
conjecture. In this conjecture, the proper action is \cite{Lehner:2016vdi}, 
\begin{align}\label{bulk}
I_{\text{WdW}}= I_{\text{bulk}}   + I_{\text{jt}} + I_{\kappa} + I_{\text{ct}},
\end{align}
which beside the bulk action, $I_{\text{bulk}}$, one needs {additional contributions from boundary terms}
\begin{align}
& I_{\text{jt}}= -2\varepsilon\int_{\rm jt}\, d^{2}\Omega\,\sqrt{q}\hspace{1mm}\mathbf{a},
\nonumber\\
&I_{\kappa} = 2 \sigma \int_{\rm WdW}\!\!\!\!\!  \ud\la\,d^{2}\Omega\,\sqrt{q}\,\kappa,
\nonumber\\
& I_{\text{ct}} = 2  \sigma \int_{{\rm WdW}}\!\!\!\!\! \ud\la \,d^{2}\Omega\,\sqrt{q}\, \Theta \log (l_{\rm ct} \Theta).
\label{IB}
\end{align}
The null joint term, $I_\text{jt}$,  should be evaluated on the intersection of null boundaries. The induced metric on the joint is $q$, and $\mathbf{a} = \log|\frac12\ell.\bar{\ell}|$ , {where} $ \ell_{\alpha } $ and $ \bar{\ell}_{\alpha} $ being the normals of intersecting null boundaries. The $ \sigma $ is a sing factor depending on position of null boundary segment with respect to the WdW patch and  $ \varepsilon $ is a sign factor depending on the position of joint respecting to the boundary segments and the WdW patch, \cite{Lehner:2016vdi}(Also see next section). The next is $I_{\kappa}$, depending on the scalar $\kappa$, which describes that how much the coordinate $\la$, which parametrises the null boundary direction fails to be an affine parameter ($\ell^\mu \nabla_{\mu} \ell_\nu=\kappa\, \ell_\nu$). The last term  is the counterterm action, $I_{\text{ct}}$, in order to ensure that the action is invariant under the reparameterization of the null boundaries, where {$l_{\text{ct}}$} is an arbitrary scale and $\Theta =\ell^\alpha\partial_\alpha \log \sqrt{q} =\partial_\la \log \sqrt{q}$ is the expansion scalar. Accordingly, the variation of complexity is given by two classes of terms:
\begin{equation}\label{var}
\delta{ \cal C}_{\text{A}}=\frac{1}{\pi}\big(\delta  I_{\text{WdW}}+I_{\delta \text{WdW}}\big),
\end{equation}
where $\delta I_{\text{WdW}}$ indicates the variation due to the change of the  background fields within the original WdW, and $I_{\delta \text{WdW}}$ is the variation {due to} the change in the shape of the WdW patch \cite{Bernamonti:2019zyy}. For $I_{\text{bulk}}$, in \cite{Bernamonti:2019zyy}, Bernamonti et. al., considered four dimensional Einstein-Hilbert gravity coupled to a free massless scalar field in order to check the {first law of complexity}. {The action for this theory is} given by
\begin{equation}
I_{\text{bulk}} = \int {\rm d}^4 x \sqrt{-g} \left[R+\frac{6}{L^2}-\frac{1}{2}\nabla^{\mu}\Phi \nabla _{\mu}\Phi
\right],
\end{equation}
where its vacuum AdS$_4$ solution is 
\begin{equation}\label{metric}
{\rm d}s^2 = \frac{L^2}{\cos^2 \rho}\left( -{\rm d }t^2+{\rm d}\rho^2+\sin ^2 \rho \,{\rm d}\Omega^2\right){ .}
\end{equation}
Here, $L$ {is} the AdS radius and the coordinate  $\rho$ has the range $[0, \pi/2]$. 
%
They studied the spherically symmetric scalar perturbations on AdS background up to second order in perturbations and showed that the variation of the holographic complexity{,} $\delta \mathcal{C}_A$, is given by
\begin{equation}\label{firstlawblackholeMyers}
\delta \mathcal{C}_A\sim\int _{\partial\text{WdW}}{\rm d} \lambda\, {\rm d}^2 \Omega \sqrt{\gamma} \hspace{.5mm}\bigg( \delta\Phi\hspace{.5mm} \partial_{\lambda}\delta\Phi \bigg){.}
\end{equation}
 It is worth noting that the changes of complexity in their analysis came entirely from the scalar field action and the gravitational contributions canceled each other. In order to investigate more their results, we study the effect of arbitrary perturbations on top of arbitrary background metric. Accordingly, we consider Einstein-Hilbert-Maxwell-Scalar theory on the D-dimensional  spacetime
\begin{equation}\label{EMP}
I_{\text{bulk}} = I_{\text{EH}}+\int {\rm d}^D x \sqrt{-g}\hspace{.5mm} \bigg(-F^{\mu \nu} F_{\nu\mu}-\frac{1}{2}\nabla^{\mu}\Phi \nabla _{\mu}\Phi-\frac{1}{2}m^2 \Phi^2
\bigg),
\end{equation}
with
\begin{equation}
I_{\text{EH}} = \int {\rm d}^D x \sqrt{-g}\hspace{.5mm}\left(R-2\Lambda\right),
\end{equation}
and the following perturbations  
\begin{align}
& g'_{\mu\nu}=g_{\mu\nu}+\delta g_{\mu\nu},\quad A'_{\mu}=A_{\mu}+\delta A_{\mu},\quad \Phi'=\Phi_0+\delta\Phi,
\label{deltas}
\end{align}
where $g_{\mu\nu},A_\mu$ and $\Phi_0$ are corresponding to  background values.
We  assume $\Phi_0=0$ and consider $\delta\Phi$ as a source of perturbations. According to this assumption, if we consider $\delta\Phi \sim \mathcal{O}(\epsilon)$, then equations of motion require perturbations  in the metric and the gauge field to  be $\mathcal{O}(\epsilon^2)$. As a result of  $\Phi_0=0$,  the on-shell action of the scalar field will become  $\mathcal{O}(\epsilon^2)$, so the whole action will also be $\mathcal{O}(\epsilon^2)$. This leads to $ \delta\mathcal{C}_{A}$ to be also $\mathcal{O}(\epsilon^2)$. In the following we will omit explicit  $\epsilon^2$ factor in our expressions, but it must be supposed.

When dealing with the variation of action in the WdW-patch, there is an important point that the null boundaries of WdW patch will no longer remain null under a generic metric perturbation, \eqref{deltas}. To make this point clear, let us suppose that before acting the perturbation, a  boundary hypersurface is characterized by a constraint $\Psi=\text{const}$. When the scalar field $\Psi$ satisfies the equation
\begin{equation}\label{NullCondition}
\nabla_{\alpha}\Psi\nabla^{\alpha}\Psi=0,
\end{equation}
this boundary becomes null. Under change of metric, $ g_{\alpha\beta}\to  g'_{\alpha\beta}=g_{\alpha\beta}+ \delta g_{\alpha\beta} $, the equation \eqref{NullCondition} changes to 
\begin{equation}\label{key}
\nabla_{\alpha}\Psi\nabla^{\alpha}\Psi-\delta g^{\alpha\beta}\nabla_{\alpha}\Psi\nabla_{\beta}\Psi = 0,
\end{equation}
which this means that  $ \Psi=\text{const}$ is no longer a null hypersurface in  the deformed geometry. Actually after a deformation, we have a  new null hypersurface $\Psi^{\prime} =\Psi+\delta\Psi = \text{const}$ which is characterized by the constraint $ g'^{\alpha\beta}\nabla_{\alpha}\Psi'\nabla_{\beta}\Psi' =0$ and this constraint is equal to
\begin{equation}\label{deformedWdW}
\nabla_{\alpha}\Psi\nabla^{\alpha}\Psi-\left(\nabla_{\alpha}\Psi\nabla_{\beta}\Psi\right)\hspace{.5mm}\delta g^{\alpha\beta}+2 \nabla_{\alpha}\Psi\nabla^{\alpha}\delta\Psi = 0.
\end{equation}
Therefore, to have $\Psi^{\prime}=\text{const}$ as a new null hypersurface, the change of boundary can be related to the change of metric as follows
\begin{equation}\label{deltaphi}
\nabla_{\alpha}\Psi\nabla^{\alpha}\delta\Psi=\frac{1}{2} (\nabla_{\alpha}\Psi\nabla_{\beta}\Psi)\hspace{.5mm}\delta g^{\alpha\beta}.
\end{equation} 
In the following, we first introduce the setup for dealing with new null hyper-surfaces and after that explore variation of on-shell action \eqref{bulk} (with \eqref{EMP}) under a general perturbation. 
\subsection{Definitions}
A hypersurface  $\mathcal N$,  characterized by $\Psi=\text{const}$, is  called a null hypersurface if and only if $\nabla_{\alpha}\Psi \nabla^{\alpha}\Psi = 0$. This feature of null boundary  indicates that the normal vector to the null surface is also tangent to it. This property is the origin of some difficulties when dealing with such hypersurfaces because, as a consequence, the induced metric becomes degenerate. As a result, constructing a projection to the null surface just from its normal is not possible. One standard remedy to this problem is to introduce an auxiliary null vector $k^{\alpha}$, which lays out of the hypersurface and therefore $\ell_{\alpha} \, k^{\alpha} \ne 0$ {when} $\ell_{\alpha}$ is the null normal to the boundary  e.g. $ \ell_{\alpha}\propto\nabla_{\alpha}\Psi $ on the boundary.\footnote{For more details about the geometry of null hypersurfaces we refer the interested reader to \cite{Poisson:2009pwt} chapter 3, see also \cite{Gourgoulhon:2005ng}. Our convention for notation almost follows \cite{Gourgoulhon:2005ng}.}  Here following \cite{Poisson:2009pwt} we chose $ \ell_{\alpha}=-\nabla_{\alpha}\Psi $ such that $ \ell_{\alpha} $ will be  future-directed when $ \Psi $
increases toward the future. We take the normalization of these null forms to be everywhere as
\begin{align}
\ell_{\alpha}\,\ell^{\alpha} =0\ ,\ k_{\alpha}\,k^{\alpha} = 0 \quad\text{and}\quad \ell_{\alpha}\,k^{\alpha}=-1\,. \label{lknormalization}
\end{align} 
With the help of $\ell_{\alpha}$ and $k_{\alpha}$, one can define the projection given by
\begin{align}
q^{\alpha}{}_{\beta} = \delta^{\alpha}{}_{\beta} + \ell^{\alpha}\,k_{\beta} + k^{\alpha}\,\ell_{\beta}.
\end{align}
This  projection is not in fact a   projection on  the null surface. Instead, it essentially projects spacetime vectors onto the co-dimension two surface $\mathcal{S}$, to which $\ell_{\alpha}$ and $k_{\alpha}$ are orthogonal.

Moreover, using the covariant differentiation of vectors $ \ell_{\alpha} $ and $ k_{\alpha} $,  and projecting them in different directions $ q^{\alpha}_{\beta} $, $ \ell^{\alpha} $ and $ k^{\alpha} $, one can define the following geometric objects:
\begin{equation}\label{BGOdef}
\begin{aligned}
&\Theta_{\alpha\beta} = q^{\sigma}{}_{\alpha}\,q^{\delta}{}_{\beta}\,\nabla_{\sigma}\ell_{\delta} \ &&, \quad \Xi_{\alpha\beta} = q^{\sigma}{}_{\alpha}\,q^{\delta}{}_{\beta}\,\nabla_{\sigma}k_{\delta}, \\
&\eta_{\alpha} = -q^{\sigma}{}_{\alpha}\,k^{\beta}\,\nabla_{\beta}\ell_{\sigma} \ &&, \quad \bar{\eta}_{\alpha} =- q^{\sigma}{}_{\alpha}\,\ell^{\beta}\,\nabla_{\beta} k_{\sigma}, \\
&\omega_{\alpha} = - q^{\sigma}{}_{\alpha}\,k^{\beta}\,\nabla_{\sigma}\ell_{\beta} = -q^{\sigma}{}_{\alpha}\,\ell^{\beta}\,\nabla_{\sigma} k_{\beta}, \\
&a_{\alpha} = q^{\sigma}{}_{\alpha}\,\ell^{\beta}\,\nabla_{\beta}\ell_{\sigma} \ &&, \quad \bar a_{\alpha} = q^{\sigma}{}_{\alpha}\,k^{\beta}\,\nabla_{\beta} k_{\sigma}, \\
&\kappa = -\ell^{\alpha}\,k^{\beta}\,\nabla_{\alpha}\ell_{\beta} = \ell^{\alpha}\,\ell^{\beta}\,\nabla_{\alpha} k_{\beta} \ &&, \quad \bar\kappa = k^{\alpha}\,\ell^{\beta}\,\nabla_{\alpha} k_{\beta} = -k^{\alpha}\,k^{\beta}\,\nabla_{\alpha}\ell_{\beta},
\end{aligned}
\end{equation}
where $\Theta_{\alpha\beta}$ and $\Xi_{\alpha\beta}$ are extrinsic curvatures of  $\mathcal{S}$, while $\omega_{\alpha}$, $\eta_{\alpha}$ and $\bar{\eta}_{\alpha}$ are twists. In addition, $a_{\alpha}$ and $\bar a_{\alpha}$ are tangent accelerations of $\ell^{\alpha}$ and $k^{\alpha}$ to $ \mathcal S$, respectively. Moreover, $\kappa$ and $\bar\kappa$ are in-affinity parameters. These geometric objects can be seen compactly in the following relations 
\begin{align}
\nabla_{\alpha} \ell_{\beta} = \Theta_{\alpha\beta} + \omega_{\alpha}\,\ell_{\beta} + \ell_{\alpha}\,\eta_{\beta} - k_{\alpha}\,a_{\beta} - \kappa\,k_{\alpha}\,\ell_{\beta} - \bar{\kappa}\,\ell_{\alpha}\,\ell_{\beta}, \label{delldecomposition}\\
\nabla_{\alpha} k_{\beta} = \Xi_{\alpha\beta} - \omega_{\alpha}\,k_{\beta} + k_{\alpha}\,\bar{\eta}_{\beta} - \ell_{\alpha}\,\bar a_{\beta} + \kappa\,k_{\alpha}\,k_{\beta} + \bar{\kappa}\,\ell_{\alpha}\,k_{\beta}, \label{delkdecomposition}
\end{align}
which are actually  generalizations of the relation $ \nabla_{\alpha}n_{\beta}=-K_{\alpha\beta}+n_{\alpha} n_{\beta} $ to  the case in hand,  where  decomposition has been done with two null vectors.

\subsection{Variations and their decompositions}
In the directions $ q^{\alpha}_{\beta} $, $ \ell^{\alpha} $ and $ k^{\alpha} $,  the general variations in metric, $ \delta g_{\mu\nu} $,  can  be decomposed into a tensor $\deltabar q_{\alpha\beta}$, two vectors $(\deltabar u_{1\alpha},\deltabar u_{2\alpha})$, and three scalars $(\deltabar \mu_{1},\deltabar \mu_{2},\deltabar \mu_{3})$ as follows
\begin{align}\label{perdefs}
&\deltabar q_{\alpha\beta} = q^{\sigma}{}_{\alpha}\,q^{\delta}{}_{\beta}\,\delta g_{\sigma \delta} \nonumber\\
&\deltabar u_{1\alpha} =- q^{\beta}{}_{\alpha}\,\ell^{\sigma}\,\delta g_{\beta \sigma} \quad,\quad
\deltabar u_{2\alpha} =- q^{\beta}{}_{\alpha}\,k^{\sigma}\,\delta g_{\beta \sigma}\quad\nonumber\\
&\deltabar \mu_{1} =\ell^{\alpha}\,\ell^{\beta}\,\delta g_{\alpha\beta} \qquad\ \,,
\quad\deltabar \mu_{2} =k^{\alpha}\,k^{\beta}\,\delta g_{\alpha\beta} \quad,\quad
\deltabar \mu_{3} =\ell^{\alpha}\,k^{\beta}\,\delta g_{\alpha\beta}.
\end{align}
Thus we have
\begin{align}
\delta g_{\alpha\beta} &= \deltabar q_{\alpha\beta} + 2 k_{(\alpha} \deltabar u_{1\beta)} + 2\ell_{(\alpha} \deltabar u_{2\beta)} + k_{\alpha} k_{\beta} \deltabar \mu_1 + \ell_{\alpha}\ell_{\beta} \deltabar \mu_2 +2 \ell_{(\alpha} k_{\beta)}\deltabar \mu_3 \,.\label{deltag}
\end{align}
Here, following  the notation of \cite{Lehner:2016vdi}, we  have introduced
$ \deltabar $  to show that these quantities are not  necessarily  variation of some function  (i.e. $ \deltabar q_{\alpha\beta}\neq\delta(q_{\alpha\beta}) $). 
We also impose that variation keeps the normalization conditions of the frame forms $ \ell_{\alpha} $ and $ k_{\alpha} $ unchanged. Therefore,  from the normalization conditions \eqref{lknormalization} together with using \eqref{deltag}, we get 
\begin{align}\label{perlk}
&0=\delta (\ell_{\alpha}\,\ell^{\alpha}) = 2 \ell^{\alpha}\delta \ell_{\alpha} - \ell^{\alpha} \ell^{\beta} \delta g_{\alpha\beta} = 2 \ell^{\alpha}\delta \ell_{\alpha} - \deltabar \mu_1 \,,\nonumber\\
&0=\delta (k_{\alpha}\,k^{\alpha}) = 2\,k^{\alpha}\,\delta k_{\alpha} - k^{\alpha}\,k^{\beta}\,\delta g_{\alpha\beta}= 2\,k^{\alpha}\,\delta k_{\alpha} - \deltabar \mu_2 \, { ,}\nonumber\\
&0=\delta (\ell_{\alpha}\,k^{\alpha}) = \ell^{\alpha}\,\delta k_{\alpha} + k^{\alpha}\,\delta \ell_{\alpha} -\ell^{\alpha}\,k^{\beta}\,\delta g_{\alpha\beta}= \ell^{\alpha}\,\delta k_{\alpha} + k^{\alpha}\,\delta \ell_{\alpha} - \deltabar \mu_3 \,.
\end{align}
By assuming that $\ell_{\alpha} $ and $ k_{\alpha} $ remain orthogonal to $ \mathcal{S} $, we can solve   the above three equations  for $\delta\ell_{\alpha} $ and $\delta k_{\alpha} $  to get
\begin{align}
\delta \ell_{\alpha} &=  \ell_{\alpha} \hspace{1mm}\deltabar\beta-\frac12\,k_{\alpha}\hspace{.5mm}\deltabar \mu_1 \label{deltal},\\
\delta k_{\alpha} &= -k_{\alpha}\hspace{.5mm}(\deltabar\beta+\deltabar \mu_3 )  -\frac12\,\ell_{\alpha} \label{deltak}\,\deltabar \mu_2,
\end{align}
where $\deltabar\beta$ is an arbitrary function which can not be fixed using relations \eqref{perlk}. In fact it is related to the rescaling  freedom in definitions of $\ell_{\alpha }$ and $k_\alpha$. Now, by using $\ell_\alpha=-\nabla_{\alpha}\Psi$ on the null hypersurface; we have  $\delta\ell_\alpha=-\nabla_{\alpha}\delta\Psi$. Thus, by using \eqref{deltal}, we get
\begin{align}\label{mubeta}
\deltabar \mu_1=-2 \ell^\alpha\nabla_{\alpha}\delta\Psi,\hspace{1cm}\deltabar \beta=k^\alpha\nabla_{\alpha}\delta\Psi. 
\end{align}
This means that just the projection of $\nabla_{\alpha}\delta\Psi$  along $\ell_\alpha$ is fixed by metric perturbations and projection along $k_\alpha$, \textit{i.e.} $ k^\alpha\nabla_{\alpha}\delta\Psi  $, will remain arbitrary. As it will be shown, this arbitrariness, in general, gives rise to some ambiguities in variation of holographic complexity. However, if we assume that $ \delta\Psi $ be function of coordinates on the null hypersurface, then $ \deltabar\beta $  vanishes. To see this, we note that we can adopt a local coordinate in spacetime using $ \Psi $, $ \la $ and $ Y^a $, where $ Y^a $ are coordinates on codimension two space $ \mathcal{S} $,(see \textit{e.g.} \cite{Poisson:2009pwt}). Then it follows that $ \ell^\alpha\partial_\alpha\propto\partial_\la $ and $ k^\alpha\partial_\alpha\propto\partial_\Psi $. Therefore of $ \delta\Psi=\delta\Psi(\la,Y^a) $, then $ \deltabar\beta=0 $. 
In what follows we keep $ \deltabar\beta $ in our calculation, to track the ambiguities in the variation of holographic complexity, however finally in the presented examples we set $ \deltabar\beta=0 $ according to the above discussion. 
\subsection{Variation of bulk term}
By varying  the action \eqref{EMP} under  perturbations \eqref{deltas} and using equations of motion and Stokes theorem we get
\begin{equation}\label{deltaII}
\delta I_{\text{bulk}}= \delta I_{\text{EH}}+{\rm sign}(\mathcal{N})\int_{\mathcal N} \ud^{D-2}\Omega\ud\la \,  \sqrt q\left(-4\ell^{\alpha}F_{\alpha\beta}\delta A^{\beta}-\frac12(\ell^\alpha\nabla_{\alpha }\delta\Phi)\delta\Phi\right),
\end{equation}
where we have called each null segments in the boundary by $ \mathcal{N} $.  Because $ \ell_{\alpha } $ is always chosen to be future directed, the  $ {\rm sign}(\mathcal{N}) =1(-1)$  indicate if the boundary is placed in the   future (past) of spacetime volume. Moreover, $\delta I_{\text{EH}}$ is given by
\begin{equation}\label{deltaEH}
\delta I_{EH}={\rm sign}(\mathcal{N})\int_{\mathcal N} \ud^{D-2}\Omega\ud\la \,  \sqrt q\left(\ell^\alpha\,\nabla^\beta \delta g_{\alpha\beta} - \ell^\alpha\,\nabla_\alpha\delta g^\beta{}_\beta\right).
\end{equation}
Now, using the expression \eqref{deltag} for $\delta g_{\alpha\beta}$ we have
\begin{align}
&\ell^\alpha\,\nabla^\beta \delta g_{\alpha\beta} - \ell^\alpha\,\nabla_\alpha\delta g^\beta{}_\beta=\nonumber\\
&\hspace{.5cm}-\ell^{\alpha}\nabla_{\alpha} \deltabar q^{\beta}{}_{\beta} - \nabla_{\alpha} \deltabar u_1^{\alpha} - k^{\alpha}\,\nabla_{\alpha}\deltabar \mu_1 +\ell^{\alpha}\,\nabla_{\alpha}\deltabar \mu_3 \nonumber\\
&\hspace{.5cm}+ \bigg(-\deltabar q_{\alpha\beta}-2\,k_{(\alpha}\,\deltabar u_{1\beta)} - \,\ell_{\alpha} \, \deltabar u_{2\beta}  -k_{\alpha}\,k_{\beta}\,\deltabar \mu_1 - (\ell_{\alpha}\,k_{\beta} + g_{\alpha\beta})\deltabar \mu_3 \bigg)\nabla^{\alpha}\ell^{\beta} \nonumber\\
&\hspace{.5cm}- \deltabar \mu_1\,\nabla_{\alpha} k^{\alpha}. \label{Pdeldeltag1}
\end{align}
According to \eqref{perdefs}, in obtaining the above result we have used the relations
\begin{align}\label{Iden1}
&\ell^{\beta}\,\nabla_{\alpha} (\deltabar u_{1\beta})= - \deltabar u_1{}^{\beta}\,\nabla_{\alpha} \ell_{\beta}, &&\ell^{\beta}\,\nabla_{\alpha} \deltabar q_{\beta\mu}=-\deltabar q_{\beta\mu}\nabla_{\alpha}\ell^{\beta},\nonumber\\
&\ell^{\beta}\,\nabla_{\alpha} k_{\beta}= - k^{\beta}\,\nabla_{\alpha} \ell_{\beta},
\end{align} 
and similar relations for $\deltabar u_2{}^{\beta}$. These are  direct consequence of normalization conditions \eqref{lknormalization} and definitions \eqref{perdefs}. Now using the expressions \eqref{delldecomposition} and \eqref{delkdecomposition} for $\nabla_{\alpha}\ell_\beta$ and $\nabla_{\alpha}k_\beta$ we get 
\begin{align}
&\hspace{.1cm}\ell^\alpha\,\nabla^\beta \delta g_{\alpha\beta} - \ell^\alpha\,\nabla_\alpha\delta g^\beta{}_\beta=\nonumber\\
&\hspace{1cm}-\mathit{a}^{\alpha } \hspace{.5mm}\deltabar \mathit{u}_{2\alpha } + 2 \bar{\kappa} \hspace{1mm}\deltabar \mu_{1} +  \left(2\eta_{\alpha }+\bar{\eta}_{\alpha }+\omega_{\alpha}\right)\hspace{.5mm}\deltabar \mathit{u}_{1}{}^{\alpha } -  \Theta^{\alpha \beta } \hspace{.5mm}\deltabar q_{\alpha \beta } -   \Theta\hspace{.5mm}\deltabar \mu_{3} - \Xi\hspace{.5mm}\deltabar \mu_{1} \nonumber\\
&\hspace{1cm} -  \ell^{\alpha } \hspace{.5mm}\nabla_{\alpha }\deltabar q^{\beta }{}_{\beta }-  \mathcal{D}_{\alpha }\deltabar \mathit{u}_{1}{}^{\alpha } -  k^{\alpha } \nabla_{\alpha }\deltabar \mu_{1} + \ell^{\alpha } \nabla_{\alpha }\deltabar \mu_{3}\label{Pdeldeltag2},
\end{align}
 where $\mathcal{D}_{\alpha}$ is the covariant derivative on the co-dimension two surface $\mathcal{S}$ orthogonal to $\ell_{\alpha }$ and $k_{\alpha}$, and in the above relation it is defined by $\mathcal{D}_{\alpha }\deltabar \mathit{u}_{1}{}^{\alpha }=q^{\alpha\beta}\nabla_{\alpha }\deltabar \mathit{u}_{1\beta }$.
As it is shown in appendix \ref{dtheta} and \ref{dkappa}, by going through similar procedure, we can find the following relation for $\delta\Theta$ and $\delta\kappa$:
\begin{align}\label{deltatheta}
&\delta\Theta= \Theta\hspace{.5mm}\deltabar \mu_{3} + \Theta\hspace{.5mm}\deltabar\beta + \frac{1}{2}  \Xi\hspace{.5mm}\deltabar \mu_{1}  + \frac{1}{2} \ell^{\alpha} \nabla_{\alpha}\deltabar q^{\beta}{}_{\beta} +  \mathcal{D}_{\alpha }\deltabar \mathit{u}_{1}{}^{\alpha }-   (\omega_{\alpha }+\bar{\eta}_{\alpha })\hspace{.5mm}\deltabar \mathit{u}_{1}{}^{\alpha }+\mathit{a}^{\alpha }\hspace{.5mm} \deltabar \mathit{u}_{2\alpha },
\nonumber\\
&
\delta\kappa=- \mathit{a}^{\alpha } \deltabar \mathit{u}_{2\alpha } -  \frac{1}{2} \bar{\kappa} \hspace{.5mm}\deltabar \mu_{1}  + \kappa\left( \deltabar \beta+ \deltabar \mu_{3}\right) +  \left(\omega_{\alpha }-\eta_{\alpha }\right)\deltabar \mathit{u}_{1}{}^{\alpha } + \ell^{\alpha } \nabla_{\alpha }\deltabar \beta + \frac{1}{2} \mathit{k}^{\alpha } \nabla_{\alpha }\deltabar \mu_{1}.
\nonumber\\
\end{align}
Solving the first equation in \eqref{deltatheta} for $\ell^{\alpha} \nabla_{\alpha}\deltabar q^{\beta}{}_{\beta}$ and second one for for $k^{\alpha} \nabla_{\alpha}\deltabar\mu_{1}$  and substituting the results in \eqref{Pdeldeltag2}, we find
\begin{align}\label{Pdeldeltag3}
&\ell^\alpha\,\nabla^\beta \delta g_{\alpha\beta} - \ell^\alpha\,\nabla_\alpha\delta g^\beta{}_\beta 
\nonumber\\
&\hspace{.5cm}=- 2 \delta \Theta- 2 \delta \kappa-  \Theta^{\alpha \beta }\hspace{.5mm}\deltabar q_{\alpha \beta } +\left(\omega^{\alpha }-\bar{\eta}^{\alpha }\right)\deltabar \mathit{u}_{1\alpha }+  \bar{\kappa} \hspace{.5mm}\deltabar \mu_{1} \nonumber\\
&\hspace{.5cm}+ 2\kappa\left( \deltabar \beta+ \deltabar \mu_{3}\right)   + \Theta\left(\deltabar \mu_{3} + 2 \deltabar\beta\right) +2  \ell^{\alpha } \nabla_{\alpha }\deltabar\beta + \ell^{\alpha } \nabla_{\alpha }\deltabar \mu_{3} +\mathcal{D}_{\alpha}\deltabar \mathit{u}_{1}{}^{\alpha}.
\end{align}
Now, this expression has to be integrated on the null boundary according to \eqref{deltaEH}. It is worth noting that after integration the last term vanishes since it is a total derivative and we suppose that the surface $\mathcal{S}$ orthogonal to the null directions is compact. Moreover, since, {$\ell_{\alpha}=-\nabla_{\alpha }\Psi$}, by using the symmetric property of $\nabla_{\alpha }\ell_{\beta}$, one can see that $\omega_{\alpha }=-\bar\eta_{\alpha }$. Furthermore, to manipulate some other terms we use the following rule for integration by parts on null surface
\begin{equation}\label{bypart}
\int_{\mathcal N}\!\!\! \ud^{D-2}\Omega\ud\la \,  \sqrt q\, \ell^\alpha\,\nabla_\alpha\boldsymbol\phi = - \int_{\mathcal N}\!\!\! \ud^{D-2}\Omega\ud\la \,  \sqrt q\ \Theta\ \boldsymbol\phi\, +{\rm sign}(\partial\mathcal{N}) \int_{\partial\mathcal N}\!\!\! \ud^{D-2}\Omega \,  \sqrt q\,\boldsymbol\phi,
\end{equation}
for every scalar $\boldsymbol\phi$. The last term is lying on the boundary of a null surface, or at the joint of intersecting null surfaces which is denoted by $ \partial\mathcal{N} $ and the  $ {\rm sign}(\partial\mathcal{N}) =1(-1)$  indicate if the joint $ \partial\mathcal{N} $ is placed in the future (past) of the boundary segment.
Accordingly, by using the relations \eqref{bypart} and \eqref{Pdeldeltag3}, the variation of bulk term \eqref{deltaEH} becomes
\begin{align}
&\hspace{-1cm}\delta I_{EH} = -2\sigma\int_{\mathcal N} \ud^{D-2}\Omega\ud\la \, \sqrt q \hspace{.5mm}\delta\hspace{-.5mm}\left( \Theta+\kappa\right)\nonumber\\
&\hspace{.4cm}-\sigma\int_{\mathcal N} \ud^{D-2}\Omega\ud\la \, \sqrt q \left( \Theta^{\alpha \beta }\hspace{.5mm}\deltabar q_{\alpha \beta } 
+2\bar{\eta}^{\alpha } \hspace{.5mm}\deltabar \mathit{u}_{1\alpha } -  \bar{\kappa} \hspace{.5mm}\deltabar \mu_{1} - 2\kappa\left( \deltabar \beta+ \deltabar \mu_{3}\right)\right)
\nonumber\\
&\hspace{.4cm}+\varepsilon\int_{\partial\mathcal N} \ud^{D-2}\Omega \sqrt q\,\left(\deltabar \mu_{3} + 2 \deltabar\beta\right) \label{surfaceterm}.
\end{align}
Where we have defined $ \sigma={\rm sign}(\mathcal{N}) $ and $ \varepsilon={\rm sign}(\mathcal{N}){\rm sign}(\partial\mathcal{N}) $. 
 To further study the last term one needs to consider the contribution from another null surface  defined by $\bar{\Psi}=\text{const}$, with normal $\bar{\ell}_{\alpha }=-\nabla_{\alpha }\bar{\Psi}$. At the joint, we have $\ell_{\alpha }\bar{\ell}^{\alpha}=e^P$ for some scalar $P$. As a result
\begin{equation}\label{lbkb}
\bar{\ell}_{\alpha}=-e^P k_{\alpha},\quad\ \text{and}\quad \bar{k}_{\alpha}=-e^{-P} \ell_{\alpha}.
\end{equation}
Corresponding to the relations \eqref{deltal}, with similar analysis we find
\begin{equation}\label{deltalbar}
\delta \bar{\ell}_{\alpha} =  \bar{\ell}_{\alpha} \hspace{1mm}\deltabar\bar{\beta}-\frac12\,\bar{k}_{\alpha}\hspace{.5mm}\deltabar \bar{\mu}_1.
\end{equation}
Now, by varying both sides of  $\ell_{\alpha }\bar{\ell}^{\alpha}=e^P$ and using \cref{perdefs,deltal,deltalbar,lbkb}, one can  find
\begin{equation}
\delta P=\deltabar\mu_{3}+\deltabar\beta+\deltabar\bar{\beta}.
\end{equation}
Summing  the joint contribution of two neighboring boundaries, we get\footnote{Note that, by using \eqref{lbkb}, we have  $\deltabar\bar{\mu}_{3}=\delta g_{\alpha\beta}\bar{\ell}^\alpha \bar{k}^{\beta}=\delta g_{\alpha\beta}\ell^\alpha k^{\beta}=\deltabar\mu_{3}$. Also one can see that always $ \bar{\varepsilon} =\varepsilon$, because for intersection segments there is two possibility, either both boundaries and joints are at future (past) or both boundaries and joint are opposite. see ref \cite{Lehner:2016vdi} }:
\begin{align}
(\deltabar\mu_{3}+2\deltabar\beta)+(\deltabar\mu_{3}+2\deltabar\bar{\beta})=2\delta P.
\end{align} 
Having this fact, the expression \eqref{surfaceterm} recast into {the} new form 
\begin{align}
&\hspace{-1cm}\delta I_{EH} =
-2\sigma\int_{\mathcal N} \ud^{D-2}\Omega\ud\la \,\sqrt q  \hspace{.5mm}\delta\hspace{-.5mm}\left( \Theta+\kappa\right) \nonumber\\
&\hspace{.4cm}-\sigma\int_{\mathcal N} \ud^{D-2}\Omega\ud\la \sqrt q\bigg(\Theta^{\alpha \beta }\hspace{.5mm}\deltabar q_{\alpha \beta }+2\bar{\eta}^{\alpha } \hspace{.5mm}\deltabar \mathit{u}_{1\alpha } - \bar{\kappa} \hspace{.5mm}\deltabar \mu_{1}- 2\kappa\left( \deltabar \beta+ \deltabar \mu_{3}\right)\bigg)
\nonumber\\
&\hspace{.4cm}
+2 \varepsilon\int_{\partial\mathcal N} \ud^{D-2}\Omega\sqrt{q}\hspace{.5mm}\delta P. \label{surfaceterm11}
\end{align}
Moreover, in general, one can add a constant to $P$. This constant can be fixed by imposing additivity of action according to \cite{Lehner:2016vdi} and it is found to  be $ -\log(2) $. This observation imply that, with  $\mathbf{a}=P-\log(2)=\log(\frac12|\ell\cdot\ell'|)$,  \eqref{surfaceterm11} is equivalent to
\begin{align}
&\hspace{-.5cm}\delta I_{EH} = -2\sigma\int_{\mathcal N} \ud^{D-2}\Omega\ud\la \,\sqrt q  \hspace{.5mm}\delta\hspace{-.5mm}\left( \Theta+\kappa\right)+2\varepsilon \int_{\partial\mathcal N} \ud^{D-2}\Omega\sqrt{q}\hspace{.5mm}\delta \mathbf{a} \nonumber\\
&\hspace{.9cm}-\sigma \int_{\mathcal N} \ud^{D-2}\Omega\ud\la \sqrt q\bigg(\Theta^{\alpha \beta }\hspace{.5mm}\deltabar q_{\alpha \beta }+2\bar{\eta}^{\alpha } \hspace{.5mm}\deltabar \mathit{u}_{1\alpha } - \bar{\kappa} \hspace{.5mm}\deltabar \mu_{1}- 2\kappa\left( \deltabar \beta+ \deltabar \mu_{3}\right)\bigg). \label{surfaceterm1}
\end{align}
Substituting back the above result in (\ref{deltaII}) gives
\begin{align}
&\delta I_{\text{bulk}}=  -2\sigma\int_{\mathcal N} \ud^{D-2}\Omega\ud\la \,\sqrt q  \hspace{.5mm}\delta\hspace{-.5mm}\left( \Theta+\kappa\right)+2\varepsilon \int_{\partial\mathcal N} \ud^{D-2}\Omega\sqrt{q}\hspace{.5mm}\delta \mathbf{a} 
\nonumber\\
&\hspace{.9cm}- \sigma\int_{\mathcal N} \ud^{D-2}\Omega\ud\la \sqrt q\hspace{.5mm}\bigg(\Theta^{\alpha \beta }\hspace{.5mm}\deltabar q_{\alpha \beta }+2\bar{\eta}^{\alpha } \hspace{.5mm}\deltabar \mathit{u}_{1\alpha } - \bar{\kappa} \hspace{.5mm}\deltabar \mu_{1}- 2\kappa\left( \deltabar \beta+ \deltabar \mu_{3}\right)\bigg)
\nonumber\\
&\hspace{.9cm}+\sigma\int_{\mathcal N} \ud^{D-2}\Omega\ud\la \,  \sqrt q\left(-4\ell^{\alpha}F_{\alpha\beta}\delta A^{\beta}-\frac12(\ell^\alpha\nabla_{\alpha }\delta\Phi)\delta\Phi\right).
\label{deltaIII}
\end{align}
The terms in the first line of \eqref{deltaIII} are total variation terms and as we will show in the subsection.\ref{variationB} they can be canceled by variation of boundary terms same as the ones in (\ref{bulk}), therefore writing the surface terms of perturbed bulk action in this form will make it easier to find changing in holographic complexity under general perturbations.  
\subsection{Variation of boundary terms}\label{variationB}
The reason for appearance of boundary terms in (\ref{bulk}) is based on this fact that the bulk action $I_{\text{bulk}}$ must be complemented with a boundary action in order to get a well defined variational principle. The boundary action for a spacelike or timelike boundary is the well-known Gibbons-Hawking term. 
For a null case, the corresponding  boundary action was discussed   in \cite{Lehner:2016vdi,Parattu:2015gga} in which it is emphasized that
to have a well-defined variational principle one needs also to some terms on the joint of spacetime segments \cite{Hayward:1993my}. 
Based on  \cite{Lehner:2016vdi}, we consider  the following terms  on the null boundaries of WdW patch, \eqref{IB}:
\begin{equation}\label{boundary}
I_{{\rm B}}=2\sigma\int_{\mathcal N} \ud^{D-2}\Omega\ud\la\sqrt q\hspace{.5mm}\bigg(\kappa+\Theta\log(l_{ct}\Theta)\bigg)-2\varepsilon\int_{\partial\mathcal N} \ud^{D-2}\Omega\sqrt q \hspace{.5mm}\mathbf{a},
\end{equation}
where $ \ud^{D-2}\Omega $ is a volume element in co-dimension two surface $ \mathcal{S} $,  $\mathbf{a} = \log\left(\frac12|\ell\cdot\bar{\ell}|\right)$ and $\ell , \bar{\ell}$ are two null vectors for two null hypersurfaces which meet at a joint. First we consider variation of the $ \kappa $ term  as follows,
\begin{align}\label{Ikappa}
\delta I_{\kappa}&= 2\sigma\int_{\mathcal N} \ud^{D-2}\Omega\hspace{.7mm} \delta({\rm d}\lambda)\sqrt q \, \kappa +\sigma\int_{\mathcal N} \ud^{D-2}\Omega\hspace{.5mm}{\rm d}\lambda\sqrt q \ q^{\alpha\beta}\deltabar q_{\alpha\beta}\ \kappa+2\sigma\int_{\mathcal N} \ud^{D-2}\Omega \hspace{.5mm}{\rm d}\lambda\sqrt q\hspace{.5mm} \delta\kappa.
\end{align}
The first term arise from changing the boundary under variation, the second term is from variation of $ \sqrt{q} $ \footnote{Note that although in general we have $ \deltabar q_{\alpha\beta}\neq \delta q_{\alpha\beta} $, but one can easily show that always we have: $  q^{\mu}_{\alpha}q^{\nu}_{\beta}\deltabar q_{\mu\nu}= \delta q_{\alpha\beta} $ and so $  q^{\alpha\beta}\deltabar q_{\alpha\beta}= q^{\alpha\beta}\delta q_{\alpha\beta} $.} and the last term is {the} variation of $ \kappa $ itself.
To calculate $ \delta({\rm d}\lambda) $, we note that on the null boundary  we have 
\begin{equation}\label{dx1}
\ud{}x^{\mu} =\ell^{\mu}{\rm d}\lambda+e^{\mu}_a {\rm d}Y^a.
\end{equation}
Where $ e^{\mu}_a $ are {bases} on co-dimension two section {space} $ \mathcal{S} $ and  $ Y^a $ are the {coordinates} on it. 
Therefore (by contracting each side with $ k_{\mu} $)
we {have  $ {\rm d}\lambda=-k_{\alpha }{\rm d}x^\alpha $, as a result }
{\begin{equation}\label{deltalambda}
\delta({\rm d}\lambda) =-\delta k_{\alpha }\,{\rm d}x^\alpha=-{\rm d}\lambda\,(\deltabar\mu_3+\deltabar\beta),
\end{equation}
where in the last step we have used the equation \eqref{deltak} and \eqref{dx1}. Therefore the variation of $ I_{\kappa} $, \eqref{Ikappa}, is simplified to
\begin{equation}\label{Ikappa1}
\delta I_{\kappa}= 2\sigma\int_{\mathcal N} \ud^{D-2}\Omega {\rm d}\lambda\sqrt q  \left(\delta\kappa-\kappa\big(\deltabar\mu_3+\deltabar\beta-\frac12  q^{\alpha\beta}\deltabar q_{\alpha\beta}\big)\right).
\end{equation}
The next step is to consider variation of the counterterm $ \Theta\log(l_{ct}\Theta) $ in \eqref{boundary}. Its variation is as follows,
\begin{align}\label{Ict}
\delta I_{ct}&= 2\sigma\int_{\mathcal N} \ud^{D-2}\Omega \hspace{.5mm}\delta({\rm d}\lambda)\sqrt q \, \Theta\log(l_{ct}\Theta)+\sigma\int_{\mathcal N} \ud^{D-2}\Omega {\rm d}\lambda\sqrt q\hspace{.5mm} q^{\alpha\beta}\deltabar q_{\alpha\beta}\ \Theta\log(l_{ct}\Theta)\nonumber\\
&+2\sigma\int_{\mathcal N} \ud^{D-2}\Omega {\rm d}\lambda\sqrt q \bigg(\delta\Theta+ \delta\Theta\log(l_{ct}\Theta)\bigg).
\end{align}
Using \eqref{deltalambda}  in the above equation we find,
\begin{align}\label{Ict1}
\delta I_{ct}&=2\sigma\int_{\mathcal N} \ud^{D-2}\Omega\ud\la \sqrt q \bigg(\delta\Theta+ \big(\delta\Theta-\Theta \deltabar\mu_3-\Theta\hspace{.5mm}\deltabar\beta+\frac12 \Theta q^{\alpha\beta}\deltabar q_{\alpha\beta}\big)\log(l_{ct}\Theta)\bigg).
\end{align}
Finally, the variation of joint term in \eqref{boundary} is given by
\begin{equation}\label{deltajt}
\delta I_{\text{jt}} =- 2\varepsilon\int_{\partial\mathcal N} \ud^{D-2}\Omega\sqrt q \hspace{.5mm}\delta \mathbf{a}-\varepsilon\int_{\partial\mathcal N} \ud^{D-2}\Omega\sqrt{q}\hspace{.5mm} \mathbf{a}\hspace{.5mm} q_{\alpha\beta}\deltabar q^{\alpha \beta }.
\end{equation}
Now, by  using   \cref{deltaIII,Ikappa1,Ict1,deltajt}, we find the following expression for variation of on-shell action under perturbations
\begin{align}\label{deltaI}
&\delta{I}_{\rm WdW}=-\sigma \int_{\mathcal N} \ud^{D-2}\Omega\ud\la \sqrt q\hspace{.5mm} \bigg(\big(\Theta^{\alpha \beta }-\kappa\ q^{\alpha\beta}\big)\deltabar q_{\alpha \beta }+2\bar{\eta}^{\alpha } \deltabar \mathit{u}_{1\alpha } - \bar{\kappa} \deltabar \mu_{1}\bigg)\nonumber\\ &\hspace{1.6cm}
+ \sigma\int_{\mathcal N} \ud^{D-2}\Omega\ud\la \sqrt q \bigg(-4\ell^{\alpha}F_{\alpha\beta}\delta A^{\beta}-\frac12(\ell^\alpha\nabla_{\alpha }\delta\Phi)\delta\Phi\bigg)
\nonumber\\
&\hspace{1.6cm}
+2\sigma\int_{\mathcal N} \ud^{D-2}\Omega\ud\la\sqrt q \,\bigg( \delta\Theta-\Theta\hspace{.5mm}\deltabar\mu_3-\Theta\hspace{.5mm}\deltabar\beta+\frac12 \Theta q^{\alpha\beta}\deltabar q_{\alpha\beta}\bigg)\log(l_{ct}\Theta)
\nonumber\\ 
&\hspace{1.6cm}
-\varepsilon\int_{\partial\mathcal N} \ud^{D-2}\Omega\sqrt{q}\hspace{.5mm} \mathbf{a}\hspace{.5mm} q_{\alpha\beta}\deltabar q^{\alpha \beta }.
\end{align}
It is worth to remind that there is also another contribution from the bulk action which is the result of changing the boundary, i.e. $I_{\delta \text{WdW}}$ in \eqref{var}. In the following we first discuss  this contribution and then we investigate  the net holographic complexity variation under general variations.   
\subsection{Variation from displacement of the boundary}\label{secdwdw}
As we have discussed earlier, the WdW patch can be  displaced by adding a general metric perturbation, so when calculating holographic complexity we must also consider the contribution on this displacement. Let us suppose that bulk action can be integrated in some normalized direction $n^{\alpha}$. This is the case for example in Einstein-Hilbert theory in spherical configurations. Therefore, the on-shell bulk action can be evaluated as 
\begin{equation}\label{int2}
\int_{\mathcal M} d^{D}x \sqrt{g}\  \nabla_{\alpha }(n^{\alpha} \mathcal{J}),
\end{equation} 
for some function $\mathcal{J}$. Integrating this term by using  Gauss theorem, and when the boundary is specified by $\phi=\text{const}$,  yields
\begin{equation}\label{int3}
\int_{\partial\mathcal M} d^{D-1}x \sqrt{g}\  n^{\alpha}\nabla_{\alpha}\phi\  \mathcal{J}.
\end{equation} 
Now, by displacing the boundary infinitesimally,  the change in the action will be
\begin{equation}\label{dM}
I_{\delta\mathcal{M}}=\int_{\partial\mathcal M} d^{D-1}x \sqrt{g}\  n^{\alpha}\nabla_{\alpha}\delta\phi\  \mathcal{J}.
\end{equation} 
Note that in the above analysis $\phi$ is not restricted. So the boundary may be  $\phi=\text{const}$ or $\phi+\delta\phi=\text{const}$, consequently, the difference between two configurations is given by the above formula.  As an  example, consider integration of Einstein-Hilbert plus cosmological constant term in spherical background 
\begin{equation}\label{blackhole}
{\rm d}s^2=-f(r){\rm d}t^2+\frac{1}{f(r)}{\rm d}r^2+r^2 {\rm d}\Omega^2_{2}.
\end{equation}
It is easy to see that
\begin{equation}
\int {\rm d}^4 x \sqrt{-g} \left(R+\frac{6}{L^2}\right)=\int r^2\sin\theta\, dr\, dt\, d\Omega \left(\frac{6}{L^2} + \frac{2}{r^2} -  \frac{2\mathit{f} (r)}{r^2}  -  \frac{4\mathit{f'}(r)}{r} - \mathit{f''}(r)\right).
\end{equation}
In this background the integration in $r$ can be performed and we can write the integral in the form of \eqref{int2}  by 
\begin{equation}\label{JJ}
\mathcal{J}=\frac{2}{ r \sqrt{\mathit{f}(r)}} \left(1 + \frac{r^2}{L^2} -  \mathit{f}(r)-\frac{r\mathit{f'}(r)}{2} \right), 
\end{equation}
where here $n^{\alpha}$ is normal vector to $r=\text{const}$ surfaces. The form \eqref{int3} helps us to evaluate integral on surfaces other than $r=\text{const}$, such as the null one. For the null boundary according to \eqref{dM}, we have
\begin{equation}\label{dwdw}
I_{\delta{\rm WdW}}=\int_{\partial{\rm WdW}} \ud^{D-2}\Omega\ud\la \sqrt{q}\  n^{\alpha}\nabla_{\alpha}\delta\Psi\  \mathcal{J},
\end{equation} 
where $n^{\alpha}$ can be decomposed as
\begin{equation}\label{n}
n^{\alpha}=\sigma (\mathfrak{b} \ell^{\alpha }-\frac{1}{2\mathfrak{b}} k^{\alpha}),
\end{equation} 
such that $n^{\alpha}$ is normalized to $1$, and $\mathfrak{b}$ is a function that depends on the metric. In the case of the black hole metric \eqref{blackhole}, $\mathfrak{b}=\sqrt{f}/2$. Using \eqref{deltal} and \eqref{n},  the equation \eqref{dwdw} changes to 
\begin{equation}\label{displace}
I_{\delta{\rm WdW}}=\sigma\int_{\partial{\rm WdW}} \ud^{D-2}\Omega\ud\la \sqrt{q} \hspace{.5mm}\bigg(\frac12 \mathfrak{b}\ \deltabar\mu_{1}+\frac{1}{2\mathfrak{b}} \deltabar\beta\bigg) \mathcal{J}.
\end{equation}
\subsection{General variation of holographic complexity}
Finally, put all the ingredients together, \eqref{deltaI} and \eqref{displace}, the variations of holographic complexity under arbitrary perturbation is given by
\begin{align}\label{Firstlaw}
&\delta\mathcal{C}=\frac{1}{\pi}(\delta{I}_{\rm WdW}+I_{\delta{\rm WdW}})
\nonumber\\
&\hspace{.4cm}=\sum_{i}\bigg(- \frac{\sigma_i}{\pi}\int_{{\mathcal N}_i} \ud^{D-2}\Omega\ud\la \sqrt q \hspace{.5mm}\bigg(\big(\Theta^{\alpha \beta }-\kappa q^{\alpha\beta}\big)\deltabar q_{\alpha \beta }+2\bar{\eta}^{\alpha } \deltabar \mathit{u}_{1\alpha } - \big(\bar{\kappa}-\frac12 \mathfrak{b} \mathcal{J}\big) \deltabar \mu_{1}+\frac{1}{2\mathfrak{b}}  \mathcal{J}\hspace{.5mm}\deltabar\beta\bigg)\nonumber\\ &\hspace{.7cm}+\frac{\sigma_i}{\pi} \int_{{\mathcal N}_i} \ud^{D-2}\Omega\ud\la\sqrt q \hspace{.5mm}\bigg(-4\ell^{\alpha}F_{\alpha\beta}\delta A^{\beta}-\frac12(\ell^\alpha\nabla_{\alpha }\delta\Phi)\delta\Phi\bigg)
\nonumber\\ 
&\hspace{.7cm} 
+\frac{2\sigma_i}{\pi}\int_{{\mathcal N}_i} \ud^{D-2}\Omega\ud\la\sqrt q \,\bigg( \delta\Theta-\Theta\hspace{.5mm}\deltabar\mu_3-\Theta\hspace{.5mm}\deltabar\beta+\frac12 \Theta q^{\alpha\beta}\deltabar q_{\alpha\beta}\bigg)\log(l_{ct}\Theta)
\nonumber\\
&\hspace{.7cm} -\frac{\varepsilon_i}{\pi}\int_{\partial{{\mathcal N}_i}} \ud^{D-2}\Omega\sqrt{q}\hspace{.5mm} \mathbf{a}\hspace{.5mm} q_{\alpha\beta}\deltabar q^{\alpha \beta }\bigg).
\end{align}
Where the sum is over each piece of the boundary including boundary segments and the joints.  The above expression {indicates}  that in general {the} variation of holographic complexity depends on the arbitrary parameter $ l_{ct} $, as well as the arbitrary function $ \deltabar\beta $. So the variation {of} holographic complexity in general {suffers} from some ambiguities. However, as we discussed earlier, the dependence in $ \deltabar\beta $ can be canceled {using  the  assumption} that the deformation function {$ \delta\Psi $} just depends on the coordinates on the null boundary, namely {$ \delta\Psi(\lambda,Y^a)$}, then it follows that $ \deltabar\beta=k^\alpha\nabla_{\alpha }\delta{\Psi}=0 $. The dependence on $ l_{ct} $ can be canceled in some special examples for the backgrounds and perturbations, as we will see in the next section.

In the following section  we will calculate explicitly \eqref{Firstlaw} for some backgrounds and perturbations. But before that it is worth noting that in above we have assumed that the gravitational perturbations result from a fluctuation in a scalar field, but the essence  of calculation is general and can be extended to any type of perturbations. For example, we can consider gravitational perturbation independently, without considering it as a result of backreaction of other fields. For example, we can obtain variations in complexity for two metrics $ g_{\mu\nu} $ and $ g_{\mu\nu}+\epsilon\delta g_{\mu\nu} $, both as solutions of Einstein {equations}.  In that case, the result will be the same as \eqref{Firstlaw} without the matter perturbations in the left hand side and the order of perturbations will be $ \epsilon $ instead of $ \epsilon^2 $.
\section{Examples}\label{Exapmle}
\subsection{AdS with non spherical perturbations}
As stated previously, the authors of \cite{Bernamonti:2019zyy} studied perturbations around AdS space and supposed these perturbations to preserve spherical symmetry. Here we will give up this assumption and study variation of holographic complexity {under  more general perturbation}.  As an  example  we consider the following perturbations around AdS {spacetime}
\begin{equation}\label{permet1}
\delta{({\rm d}s^2)}= f(r)(\deltabar\mu_{3}+\frac{1}{2}\deltabar\mu_{1}f(r)){{\rm d}t^2}- \frac{1}{f(r)}(\deltabar\mu_{3}-\frac{1}{2}\deltabar\mu_{1}f(r)){{\rm d}r^2}+{\deltabar\rho}\ (r^2{\rm d}\theta^2+r^2\sin^2\theta{\rm d}\phi^2),
\end{equation}
where $\deltabar\mu_{1},\deltabar\mu_{3}$ and   $ \deltabar\rho $  are  arbitrary functions that must satisfy linearized equations of motion. This form of perturbations are chosen such that $ \deltabar\mu_{1}=\ell^\alpha\ell^{\beta}\delta g_{\alpha\beta} $ and  $ \deltabar\mu_{3}=k^\alpha\ell^{\beta}\delta g_{\alpha\beta} $, furthermore $ \deltabar\rho $  is a component of  $\deltabar q_{\alpha\beta}$ as a perturbation on the sphere.  In order to find {the variation} of holographic complexity under these perturbations, according to general rule \eqref{Firstlaw}, we must find some geometric quantities for the background such as $ \Theta_{\alpha \beta },\bar{\eta}_{\alpha} $ and $ \bar{\kappa} $, the integration of on-shell bulk Lagrangian on null boundaries $\mathcal{J}$, and project perturbations on  directions tangential to null boundaries using $ q^{\alpha\beta} $ and $ \ell^{\alpha} $. To compute those geometric quantities one needs the one form $ \ell_{\alpha} $ and the vector $ k^{\alpha} $. For AdS space in global coordinates{, which} WdW patch boundaries are given by $ t\pm L \tan ^{-1}\left(\frac{r}{L}\right)=\text{const}$, these tensors are given by: 
\begin{align}\label{geoAdS}
&\ell_{\alpha }{\rm d}x^{\alpha}=-{\rm d}t\mp\frac{1}{1+\frac{r^2}{L^2}}{\rm d}r\nonumber\\
&k^{\alpha }\partial_{\alpha}=\frac12\partial_t\pm\frac12 (1+\frac{r^2}{L^2})\partial_r.
\end{align} 
Using these, we can find easily $ \Theta_{\alpha \beta }\ud\,x^\alpha\ud\,x^{\beta}=\pm r(\ud\theta^2+\sin^2\theta\ud\phi^2) $, and $ \bar{\kappa}=\pm\frac{r}{L^2} $. The integration of $ R-2\Lambda=\frac{-6}{L^2} $ according to \eqref{JJ} leads to $ \mathcal{J}= \mp\tfrac{2r}{L^2}(1+\tfrac{r^2}{L^2})^{-1/2}$. One can also see easily that $ \bar{\eta}_{\alpha }=0 $. Moreover, according to the discussion after \eqref{Firstlaw}, we  choose $ \deltabar\beta=0 $, also by doing straightforward calculation we find the following expression for $ \delta\Theta $
\begin{equation}\label{deltathetaS}
\delta\Theta=\tfrac{2}{r}\deltabar\mu_3-\partial_{r}{\deltabar\rho}+\tfrac{1}{1+\tfrac{r^2}{L^2}}\partial_{t}{\deltabar\rho}.
\end{equation} 
Substituting  these perturbations in \eqref{Firstlaw}, and  AdS for the  background metric, and using a by part integration we find: 
\begin{align}\label{firstlawads}
\delta \mathcal{C}_{A} =&\sum_{i}\bigg( \frac{\sigma_i}{\pi}\int_{\mathcal{N}_i} {\rm d}^2\Omega \ud\lambda \sqrt{q}\left(  \frac{4}{r}\log(\frac{2l_{ct}}{r})\hspace{.5mm}{\deltabar\rho}-\frac12\ell^{\alpha} \delta\Phi\nabla_{\alpha}\delta\Phi\right)\nonumber\\
& \hspace{.7cm}+\frac{2\varepsilon_i}{\pi}\int_{\partial\mathcal{N}_i} {\rm d}^2\Omega\sqrt{q}\  \log\big(\frac{2l_{ct}}{r}(1+\frac{r^2}{L^2})\big)\hspace{.5mm}{\deltabar\rho}\bigg).
\end{align}
As a result, if {$\deltabar\rho=0$}, which is the case when perturbation preserve spherical symmetry, the above expression leads to the results of \cite{Bernamonti:2019zyy}, and just the contribution of scalar field remains. It is worth noting that here we have found firstly a general expression for change of holographic complexity, \eqref{Firstlaw}, and then specialize the background and perturbations. The other distinction between our work and  \cite{Bernamonti:2019zyy} is that we considered the contributions of  the joint terms and these terms automatically cancel the surface integral (i.e. total variation term in \eqref{deltaIII}) when doing integration by parts. In contrast, to remove those surface integrals, the authors of   \cite{Bernamonti:2019zyy} used the known asymptotic fall-off for the perturbations.  Other important aspect of the result \eqref{firstlawads} is that in general we see that the gravitational perturbations appear  in the variation of holographic complexity even in AdS background, when we do not restrict ourselves to spherical preserving perturbations. 
\subsection{Charged AdS black hole}\label{ChargeAdS}
For second example we consider the charged AdS black hole in four dimensional spacetimes. The metric is given by \eqref{blackhole} where the blackening factor is given by
\begin{equation}\label{fsol}
f(r) = 1+\frac{r^2}{L^2}-\frac{2M}{r}+\frac{Q^2}{r^2}.
\end{equation}
We perturb {the solution} \eqref{blackhole} by turning on the scalar field on top of that and let it to backreact. Here we assume that the backreaction respects the spherical symmetry, therefore the perturbation  are the same as \eqref{permet1} with {$\deltabar\rho=0$} and $ f(r) $ in \eqref{fsol}. 
{As a result  for these perturbation  $\deltabar{q}_{\alpha\beta}=\deltabar u_{1\alpha}=0$}. 
The one form $\ell_{\alpha }$ and vector $k^{\alpha }$ for the background are give by
\begin{align}\label{lkblackhole}
&\ell_{\alpha}dx^\alpha=-dt\mp\frac{1}{f(r)}dr, \nonumber\\
&k^{\alpha}\partial_\alpha=\frac{1}{2} \partial_{t}\pm\frac{f(r)}{2} \partial_{r}.
\end{align} 
According to the definition of  $\bar{\kappa}$ in \eqref{BGOdef} one can easily find $\bar{\kappa}=\pm\frac{1}{2}f'(r)$. Moreover, $\mathcal{J}$ is given by \eqref{JJ}. Substituting back them in expression \eqref{Firstlaw} we find that
\begin{align}
&\delta \mathcal{C}_{A} =\sum_{i} \frac{\sigma_i}{\pi}\int_{\mathcal{N}_i}{\rm d} \la\, {\rm d}^2 \Omega \sqrt{q}\hspace{.5mm} \bigg( \frac{M}{r^2}\deltabar\mu_{1}-4 \ell_{\beta} F^{\alpha \beta}\delta A_{\alpha}-\frac12(\ell^\alpha\nabla_{\alpha }\delta\Phi)\delta\Phi \bigg){ .}
\label{firstlawblackhole}
\end{align}
For $M=0$ and when the Maxwell field is turned off, again this result becomes  the one in \cite{Bernamonti:2019zyy}. We see  that for this black hole case also the gravitational contributions {do not} completely cancel each other in contrast to pure AdS background. Actually,  those contributions appear  proportional to the energy of spacetime. In this section we used our general formula \eqref{Firstlaw} in order get the result \eqref{firstlawblackhole}, however this result can be obtained in a more direct way as in \cite{Bernamonti:2019zyy}, \emph{i.e.}. 
direct evaluations of bulk surface terms as well boundary and joint terms. This method will be presented in appendix \ref{AppB}.
\subsection{Slow rotating AdS black hole}
In the first law of holographic complexity, \eqref{firstlawblackhole}, we have seen that the coefficient of the perturbation $\deltabar\mu_{1}$  is related to the energy of spacetime. A subsequent question {would be}  the  physical meaning of other perturbation components. In this subsection, we will show that the coefficient  of $\deltabar u_{1\alpha}$  is directly related to the angular momentum of spacetime. To examine this statement, we consider the AdS black hole with angular momentum (known as  Kerr-AdS black hole) as the background. For the sake of  simplicity in our calculations, we suppose {that the} spin of the black hole {is} small and work {up to linear order} in angular parameter  $\mathfrak{a}$. So we consider the background metric as
\begin{align}
{\rm d}s^2 &= -f(r){\rm d}t^2+\frac{1}{f(r)}{\rm d}r^2
+ 2 \mathfrak{a}  \left(f(r)-1\right)\sin ^2\theta{\rm d}t\, {\rm d}\phi+r^2\left({\rm d}\theta^2+ \sin ^2\theta \,{\rm d}\phi^2\right).
\label{Kerr}
\end{align}
Also, we assume the perturbations as 
\begin{equation}\label{per1}
\delta{(ds^2)}=  \mathfrak{a}\frac{\ c(t,r)}{r^2}\left(f(r)-1\right) {\rm d}t^2+\frac{1}{2} c(t,r) {\rm d}t {\rm d}\phi.
\end{equation}
For these perturbations, we have  $\deltabar{}q_{\alpha\beta}=\deltabar \mu_1 = 0$, and $\deltabar u_{1\alpha}{\rm d}x^\alpha=-\frac{c(t,r)}{2f(r)}{\rm d}\phi$. Moreover,  $\ell_{\alpha }$ and $k^{\alpha }$ for the background metric \eqref{Kerr} to first order  in $ \mathfrak{a} $ are the same as \eqref{lkblackhole}.
Furthermore, by definition of $\bar{\eta}_{\alpha}$ in \eqref{BGOdef}, one can see that $\bar{\eta}_{\alpha}dx^{\alpha}=\frac{3a M\sin^2\theta}{r^2}d\phi$. Substituting back all these expressions in \eqref{Firstlaw} we find that
\begin{equation}
\delta \mathcal{C}_{A} =-\sum_{i} \frac{\sigma_i}{\pi}\int_{\mathcal{N}_i} \ud^{D-2}\Omega \ud\la \sqrt{q}\left(3 \, \mathfrak{a}\,M\frac{c(t , r)}{\, r^4 f(r)}+4\ell_{\beta} F^{\alpha \beta}\delta A_{\alpha}+\frac{1}{2}(\ell^{\alpha} \nabla_{\alpha}\delta\Phi)\delta\Phi\right),
\end{equation}
where the result depends on the angular momentum of the Kerr-AdS black hole $J  = \mathfrak{a} M $. The fact that $\bar{\eta}_{\alpha}$ is related to the angular momentum of  spacetime also has been show explicitly in \cite{Jafari:2019bpw} for asymptoticly flat spacetimes.  
\section{Discussion and results}\label{discussion}
In this paper, we have generalized the first law of holographic complexity, proposed in  \cite{Bernamonti:2019zyy}, for arbitrary perturbations and general backgrounds. We observe that mass and  angular momentum are the responses for  perturbations in the direction of null geodesics.
 To be more concrete, the perturbations, which appear in (\ref{Firstlaw})  are all tangent to the null hypersurface, which following the logic of Brown and York \cite{Brown:1992br}, we can interpret their momentum conjugate on the null boundary as components of stress tensor defined on  this hypersurface.   
In other words one can interpret the first law \eqref{Firstlaw} as (considering just gravitational perturbations for the moment)
\begin{align}\label{Firstlaw1}
&\delta\mathcal{C}=\sum_{i} \frac{\sigma_i}{\pi}\int_{\mathcal{N}_i} \ud^{D-2}\Omega \ud\la \sqrt q\  T_{\text{\tiny{nBY}}}^{\alpha\beta}\delta g^{\|}_{\alpha\beta},
\end{align}
where $T_{\text{\tiny{nBY}}}$ is the counterpart of Brown-York stress tensor defined on the null boundaries, and by $  \delta g^{\|}_{\alpha\beta} $ we mean components of metric perturbations tangential to the boundary. The explicit relation of $ T_{\text{\tiny{nBY}}} $ with stress tensor of dual boundary theory is not evident up to now\footnote{We hope to address this issue in future publications}. But it is worth mention that a similar stress tensor for the null hypersurfaces is proposed in \cite{Jafari:2019bpw} in which a general double foliation has been used for description of  null hypersurfaces. This general double foliation can be used both for null and non-null boundaries. This general framework, which has been discussed in detail in \cite{Aghapour:2018icu}, help us to define an stress tensor similar to Brown-York tensor on null boundaries. In this  method, as in the Brown-York procedure, the resulting quasi local energy becomes infinite and the counterterms from reference space time has been used to make the result finite. In this paper we didn't use the double foliation, and instead {the} variation of boundary defining scalar function has been used in calculations. Furthermore, following \cite{Bernamonti:2019zyy}, we paid attention to contributions from bulk action{, when} varying the boundary. In fact, this is exactly these contributions that make the stress tensor finite without needing for counterterms. In this sense our result may be interesting from pure gravitational point of view because it introduces a new method for getting gravitational charges by definition of "quasi-local gravitational stress tensor" on the null hypersurfaces. This interpretation as stress tensor also has been discussed in the recent paper \cite{Bernamonti:2020bcf}.   

Finally, let us comment on the implications of our results for complexity in the field theory. Using Nielsen's approach to circuit complexity, one can find that the first variation of complexity takes the form of
\begin{equation}
\delta\mathcal{C}=p_a\delta x^a\quad {\rm with}\quad p_a=\frac{\partial F}{\partial\dot{x}^a},
\end{equation}
for some cost function $ F $. In \cite{Bernamonti:2019zyy}, using their result in holographic complexity, they deduce that the direction along the path $ p_a $ is probably orthogonal to the variation of the target state $ \delta x^a $, because for the background and perturbations they have considered, first order variations vanish. Our finding (see the paragraph in below \eqref{Firstlaw}) reveals  that for general target states this is not true, and in general, the first order variations of the complexity will not vanish.
\acknowledgments
We would like to thank Mohsen Alishahiha, Shan-Ming Ruan, Joan Simón and Behrad Taghavi for useful
comments and also Robert Myers for correspondences. We also thank  Mostafa Ghasemi, Hamed Zolfi, Mehregan Doroudiania, and Reza Pirmoradian for fruitful discussions on related topics.  S. Sedigheh Hashemi would like to
thank Iran Science Elites Federation for their support during this
project. Finally we would like to thank the referee for his/her constructive comments.
\appendix
\section{Some details of calculations}\label{AppA}
In this appendix we present some details for calculations that are given in section.\ref{sec:general}. 
Details of calculations can also be followed  in a supplementary Mathematica notebook{, which} the capability of abstract tensor package xAct \cite{MARTINGARCIA2008597,Nutma:2013zea} has been used for calculations.
\subsection{Variation Of $\Theta$}\label{dtheta}
From the definition for $\Theta_{\alpha\beta}$ in \eqref{BGOdef} we have {found} 
\begin{align}
\delta \Theta = q^{\alpha\beta}\,\delta \Theta_{\alpha\beta} - \Theta^{\alpha\beta}\,\deltabar q_{\alpha\beta} =q^{\alpha\beta}\,\delta (\nabla_{\alpha }\ell_{\beta}) - \Theta^{\alpha\beta}\,\deltabar q_{\alpha\beta}.
\end{align}
For the first term we get 
\begin{align}
q^{\alpha\beta}\,\delta (\nabla_{\alpha }\ell_{\beta})=q^{\alpha\beta}\,\nabla_{\alpha }\delta \ell_{\beta}-\delta\Gamma^{\rho}_{\alpha\beta}\ell_{\rho}.
\end{align}
Using relation \eqref{deltal} for $\delta\ell_\alpha$ we find
\begin{equation}
q^{\alpha\beta}\,\nabla_{\alpha }\delta \ell_{\beta}=   (q^{\alpha \beta }\nabla_{\beta }\ell_{\alpha })\hspace{.5mm}\deltabar\beta - \frac{1}{2} (q^{\alpha \beta } \nabla_{\beta}k_{\alpha })\deltabar \mu_{1} = \Theta\hspace{.5mm}\deltabar \beta-  \frac{1}{2} \Xi\hspace{.5mm}\deltabar\mu_{1}.
\end{equation}
Also by using the standard expression for $\delta\Gamma^{\rho}_{\alpha\beta}$
\begin{equation}
\delta\Gamma^{\rho}_{\alpha\beta}=\frac12g^{\rho \sigma } (\nabla_{\alpha }\delta \mathit{g}_{\beta \sigma } + \nabla_{\beta }\delta \mathit{g}_{\alpha \sigma } -  \nabla_{\sigma }\delta \mathit{g}_{\alpha \beta }),
\end{equation}
and using the relation \eqref{deltag}, definitions \eqref{BGOdef} and some identities similar to \eqref{Iden1} and after  some straightforward algebra we get
\begin{align}
&\delta\Theta=  \Theta\hspace{.5mm}\deltabar\mu_{3}+  \Theta\hspace{.5mm}\deltabar\beta + \frac{1}{2} \Xi\hspace{.5mm}\deltabar \mu_{1} + \frac{1}{2} \ell^{\alpha} \nabla_{\alpha}\deltabar q^{\beta}{}_{\beta} +  \mathcal{D}_{\alpha }\deltabar \mathit{u}_{1}{}^{\alpha }- \left(\omega_{\alpha }+\bar{\eta}_{\alpha }\right)\deltabar \mathit{u}_{1}{}^{\alpha }+\mathit{a}^{\alpha } \deltabar \mathit{u}_{2\alpha }.
\end{align}
The last  term indeed vanishes when $\ell_{\alpha}=-\nabla_{\alpha }\Psi$ as {can be seen easily}
\begin{align}
&a_{\alpha} = q^{\sigma}{}_{\alpha}\,\ell^{\beta}\,\nabla_{\beta}\ell_{\sigma}=-q^{\sigma}{}_{\alpha}\,\ell^{\beta}\,\nabla_{\beta}\nabla_{\sigma}\Psi=-q^{\sigma}{}_{\alpha}\,\ell^{\beta}\,\nabla_{\sigma}\nabla_{\beta}\Psi=q^{\sigma}{}_{\alpha}\,\ell^{\beta}\,\nabla_{\sigma}\ell_{\beta}=0.\nonumber
\end{align}
\subsection{Variation Of $\kappa$}\label{dkappa}
In order to find expression for $ \delta\kappa $ we act similar to the procedure for  $ \delta\Theta $ in previous subsection. Starting with the definition of $ \kappa $ as $ \kappa = -\ell^{\alpha}\,k^{\beta}\,\nabla_{\alpha}\ell_{\beta}  $ and varying it we get:
\begin{align}
\delta\kappa=&-g^{\alpha \sigma} \delta\ell_{\sigma} k^{\beta } \nabla_{\alpha }\ell_{\beta } -  g^{\beta \sigma} \delta k_{\sigma} \ell^{\alpha } \nabla_{\alpha }\ell_{\beta } + k^{\beta } \ell_{\sigma} \delta \mathit{g}^{\alpha \sigma} \nabla_{\alpha }\ell_{\beta } + k^{\sigma} \ell^{\alpha } \delta \mathit{g}_{\beta \sigma} \nabla_{\alpha }\ell_{\beta }\nonumber\\& + \frac{1}{2} k^{\beta } \ell^{\sigma} \ell^{\alpha } \nabla_{\alpha }\delta \mathit{g}_{\sigma\beta } + \frac{1}{2} k^{\beta } \ell^{\sigma} \ell^{\alpha } \nabla_{\beta }\delta \mathit{g}_{\sigma\alpha }- \frac{1}{2} k^{\beta } \ell^{\sigma} \ell^{\alpha } \nabla_{\sigma}\delta \mathit{g}_{\alpha \beta } -  k^{\beta } \ell^{\alpha } \nabla_{\alpha }\delta\ell_{\beta }
\end{align}
The next step is to use expressions \eqref{deltal} and \eqref{deltak} for $ \delta\ell_{\alpha } $ and $ \delta k_{\alpha} $ respectively,  also the expression \eqref{deltag} for metric variation and then using expressions \eqref{delldecomposition} and \eqref{delkdecomposition} for $ \nabla_{\alpha }\ell_{\beta } $ and  $ \nabla_{\alpha }k_{\beta } $ and after some tensor algebra we find: 

\begin{equation}
\delta\kappa=- \mathit{a}^{\alpha } \deltabar \mathit{u}_{2\alpha } -  \frac{1}{2} \bar{\kappa} \deltabar \mu_{1}  + \kappa( \deltabar \beta+ \deltabar \mu_{3}) + \left(\omega_{\alpha }-\eta_{\alpha }\right)\deltabar \mathit{u}_{1}{}^{\alpha } + \ell^{\alpha } \nabla_{\alpha }\deltabar \beta + \frac{1}{2} \mathit{k}^{\alpha } \nabla_{\alpha }\deltabar \mu_{1}
\end{equation}
\section{Another derivation of result in :\hspace{1mm}\ref{ChargeAdS}}\label{AppB}
As we have seen in \cref{ChargeAdS}, variations in holographic complexity in the case of black hole background leads to terms linearly proportional to the black hole mass. Here we follow another method to obtain this result, which means we use evaluation of bulk surface and boundary terms directly for black hole background and specific perturbations.  
Before turning on the perturbations, the null boundaries   are defined by  hypersurfaces which are determined  by
\begin{equation}\label{null}
\Psi_{\pm} =\pm t - r^*(r)-\Psi_0=0,
\end{equation}
where $r^*$  is the tortoise coordinate, $r^*(r)= \int dr/f(r)$.  Each of $ \Psi_{+} $ and $ \Psi_{-} $ can be a future or past boundary in the WdW patch, depending on the value of $ \Psi_0 $. So in total we have four boundaries two of them in future and  two in the past. There is also four joint in the intersection of these boundaries.
It can be easily checked {that for the null boundaries} \eqref{null} and background metric \eqref{blackhole} we have
$\nabla^{\alpha}\Psi_{\pm} \nabla _{\alpha}\Psi_{\pm} =0.$
However, in the perturbed geometry{,} this relation is no longer satisfied. Let us assume that the equation of null hypersurface in the perturbed geometry is given by $\Psi_{\pm}' = \Psi_{\pm}+\delta\Psi_{\pm}=\text{const}$, where $\delta\Psi_{\pm}$ is a small arbitrary function.
Accordingly, one finds the following relation between $ \delta\Psi_{\pm}$ and {the} metric perturbations
\begin{align}\label{deltapsi}
\partial_{r}\delta\Psi_{\pm}=\frac{1}{2} \deltabar\mu_{1}+\frac{1}{f(r)}\partial_{t}\delta\Psi_{\pm}{.}
\end{align}
This differential  equation can be  solved to find  $ \delta\Psi_{\pm}$  as {an} integral function of $ \deltabar\mu_1 ${, for example, by assuming  that $ \delta\Psi_{\pm}$ to be a function of $r$}. However  in what follows we do not  need integral form of $\delta\Psi_{\pm}$, and {only by using} this equation, our final result will we found in terms of metric perturbations such as $ \deltabar\mu_{1} $.  Note that to first order in perturbation by imposing boundary equation  $\Psi_{\pm} =\pm t - r^*(r)=\text{const} $, $\delta\Psi_{\pm}$ is just a function of $ r$. Therefore, working to first order in perturbations we set $\partial_{t}\delta\Psi_{\pm}=0$. On the other hand, the null  normal to the boundaries in the original and perturbed geometry  are given by $\ell_{\alpha }=-\nabla_{\alpha}\Psi_{\pm}$, and $\ell'_{\alpha }=-\nabla_{\alpha}\Psi_{\pm}'=\ell_{\alpha }+\delta\ell_{\alpha }$, where $ \delta\ell_{\alpha }=-\nabla_{\alpha }\delta\Psi_{\pm}$. 
The expression for $ \ell'_{\alpha } $ is  as follows: 
\begin{equation}\label{ellp}
\ell'_{\alpha} dx^{\alpha}=\mp{}dt-\left(\frac{1}{f(r)}+\frac12\deltabar\mu_{1}\right)dr.
\end{equation}
Here, and what follows, the  lower  (upper) sign choice corresponds to the $ \Psi_{-} $ ($\Psi_{+}$) boundary.
As we mentioned earlier, the null vector satisfies the geodesics equation $\ell^\alpha \nabla_{\alpha} \ell_\beta=\kappa\, \ell_\beta$, where $\kappa$ measures the failure of $\la$ to be an affine parameter on the null generators. Although  $\kappa$ vanishes for the background, using above definition can  found $\kappa$ for perturbed geometry as
\begin{equation}\label{kappap}
\kappa'=\delta\kappa=\pm\frac12\partial_{t}\deltabar\mu_1
\end{equation}
The expansion scalar{,} $\Theta${,} for the perturbed metric can also be calculated and  gives 
\begin{align}\label{thetarelation}
\Theta' &= \ell'^{\alpha}\partial_{\alpha}\log(\sqrt{q})=\frac{2}{r} \left(1+\deltabar\mu_{3}\right),
\end{align}
so we have 
\begin{align}
\delta \Theta = \frac{2}{r} \deltabar\mu_{3}=\Theta\, \deltabar\mu_{3}.
\label{deltaTheta}
\end{align}
This relation can easily be checked by noting that  because we assume that spherical symmetry is respected by perturbation, we have 
\begin{equation}\label{deltateta}
\delta\Theta=\delta\ell^{\alpha}\partial_{\alpha}\log(\sqrt{q}).
\end{equation}
Then using $ \sqrt{q}=r^2\sin\theta $ and 
\begin{equation}\label{deltaell1}
\delta\ell^{\alpha}\partial_{\alpha}=\mp\left(\frac12\deltabar\mu_{1}+\frac{1}{f(r)}\deltabar\mu_{3}\right)\partial_{t}+\deltabar\mu_{3}\partial_{r},
\end{equation}
the result \eqref{deltaTheta}  will be established. 
Now, {with the} above expressions, the contributions coming from the action (\ref{bulk}) can be obtained. Firstly for the $ I_\kappa $ term we have, 
\begin{equation}\label{dIk}
\delta I_{\kappa}=\pm \sigma \int_{\rm WdW}\!\!\!\!\!  \ud\la\,d^{2}\Omega\,\sqrt{q}\, \partial_{t}\deltabar\mu_1.
\end{equation}
On the other hand variation of the counterterm leads to,
\begin{align}
\delta I_{\text{ct}} &
= 2\sigma\int_{\partial\delta\text{WdW}}\!\!\!\! \ud\la\,d^{2}\Omega\,\sqrt{q}\, (\Theta_0+\delta\Theta)\log \ell_{\text{ct}}(\Theta_0+\delta\Theta) -  2\sigma\int_{\partial\text{WdW}}\!\!\!\! \ud\la\,d^{2}\Omega\,\sqrt{q}\,\Theta_0\log \ell_{\text{ct}}\Theta_0\,\nonumber\\&
=2\sigma\int_{\partial \text{WdW}}\ud\la\,{\rm d}^2 \Omega\, \sqrt{q}\, \delta \Theta
=2\sigma\int_{\partial \text{WdW}}\ud\la\,{\rm d}^2 \Omega\, \sqrt{q}\,\frac{2}{r}\deltabar\mu_{3} 
\label{dCT}.
\end{align}
The first line include both contributions from changing the boundary counterterm and changing boundary itself. From the first line to the first expression in the second line in \eqref{dCT} we need $ \delta(\ud\la)\ \Theta=-\ud\la \ \delta\Theta $, 
According to \eqref{thetarelation}, we have $ \delta\Theta=\deltabar\mu_3\Theta $. Furthermore  we have $ \delta(\ud\la) =-\deltabar\mu_{3} \ud\la$, and so we  get the desired result $ \delta(\ud\la) \Theta=-\ud\la \ \delta\Theta $. 
	The other contribution is from the joint terms as 
\begin{equation}\label{joint1}
I_{\text{jt}}=2 \varepsilon \int_{\rm jt}\,  {\rm d}^2 \Omega\,\sqrt{q}\ \mathbf{a}.
\end{equation}
 By using $\ell_{\alpha}$ and $\bar{\ell}_{\alpha}$ for normals to {the} intersecting null boundaries{,}  we {find} 
\begin{align}\label{arelation}
\delta\mathbf{a}&=\delta\log\left(\frac12|g^{\alpha\beta}\ell_{\alpha }\bar{\ell}_{\beta}|\right)=\frac{1}{\ell\cdot\bar{\ell}}\left(-\delta g_{\alpha\beta}\ell^\alpha\bar{\ell}^{\beta}+\bar{\ell}^{\alpha}\delta\ell_{\alpha}+\ell^{\alpha}\delta\bar{\ell}_{\alpha}\right)=\deltabar\mu_{3}+\frac{f(r)}{2}\deltabar\mu_{1}.
\end{align}
Where in the last step we have used   $ \ell\cdot\bar{ \ell}=\frac{2}{f(r)} $ and  $ \delta g_{\alpha\beta}\ell^\alpha\bar{\ell}^{\beta}=\frac{2\deltabar\mu_{3}}{f(r)} $.  
As a result 
\begin{equation}\label{jointt1}
\delta I_{\text{jt}}= 2\varepsilon \int_{\rm jt}\,  {\rm d}^2 \Omega\,\sqrt{q}\left(\deltabar\mu_{3}+\frac{f(r)}{2}\deltabar\mu_{1}\right).
\end{equation}
Having the expressions for variations of {the} boundary and {the} joint terms, the next step is to find  the variation of the bulk action. The contribution from gravitational part of the action is   given by
\begin{align}
\delta I_{\text{bulk}}&=\sigma\int _{{\partial \text{WdW}}}{\rm d}\la\, {\rm d}^ 2 \Omega\, \sqrt{q}\, \ell_{\beta}\left(\nabla_{\alpha}\delta g^{\alpha \beta}-\nabla^{\beta}\delta g^{\alpha}_{~\alpha}\right)\nonumber\\&=\sigma
\int_{{\partial \text{WdW}}} {\rm d}\la\, {\rm d}^ 2 \Omega \, \sqrt{q}\,\bigg( - \frac{2 \deltabar \mu_{3}}{r} +  \deltabar \mu_{1}{}\mathit{f}' (r) + \mathit{f} (r) \bigl(\frac{ \deltabar \mu_{1}{}}{r} +\frac12 \partial_r \deltabar \mu_{1}{}\bigr) + \partial_r \deltabar \mu_{3}{} \nonumber\\&\pm\frac12 \partial_t \deltabar \mu_{1}{} \mp\frac{1 }{\mathit{f} (r)}  \partial_t \deltabar \mu_{3}\bigg).
\end{align}
Noting that $ \partial_\lambda=\ell^{\alpha}\partial_{\alpha}=\mp\tfrac{1}{f(r)}\partial_t+\partial_r $, we  can rewrite the above expression as 
	\begin{align}\label{bulk1}
	\delta I_{\text{bulk}}&=\sigma
	\int_{{\partial \text{WdW}}} {\rm d}\la\, {\rm d}^ 2 \Omega \, \sqrt{q}\,\bigg(  \frac{4 \deltabar \mu_{3}}{r} \mp\partial_{t}\deltabar\mu_{1}- \tfrac12 \deltabar \mu_{1}{}\mathit{f}' (r) \bigg)\nonumber\\&+\sigma\int_{{\partial \text{WdW}}} {\rm d}\la\, {\rm d}^ 2 \Omega \ \partial_\lambda\bigg(\sqrt{q}(\deltabar\mu_{3}+\frac{f(r)}{2}\deltabar\mu_{1})\bigg).
	\end{align}
The contribution from {the} matter fields is given by
\begin{equation}\label{bultmatter}
\delta I_{\text{bulk, matter}}=\sigma\int_{\mathcal N} {\rm d}\la\, {\rm d}^2\Omega  \,  \sqrt q\,\left(-4\ell^{\alpha}\delta A^{\beta}F_{\alpha\beta}-\frac12\delta\Phi\ell^\alpha\nabla_{\alpha }\delta\Phi\right).
\end{equation}
In addition,  $I_{\delta \text{WdW}}$ can be determined by integrating the on-shell, zero{th} {order} Lagrangian density over the additional spacetime volume closed off by the perturbed WdW patch. The result leads to 
\begin{align}\label{deltawdw}
I_{\delta \text{WdW}}&=\int_{\delta \text{WdW}} {\rm d}^4x\sqrt{-g_{0}}\left(R(g_0)+\frac{6}{L^2}-F_{\alpha\beta}F^{\beta\alpha}\right)
\nonumber\\&=\int_{\delta \text{WdW}} {\rm d}^4x\, \sqrt{g_0}\,\left(- \frac{6}{L^2} - \frac{2 Q^2}{r^4}\right),
\end{align}
where $R(g_0)= -12/L^2$ for charged AdS black hole (\ref{blackhole}) and the solution for $A_{\alpha}$ has been used, where the only non-vanishing component of $ F_{\alpha\beta} $ is $ F_{tr}=\frac{Q}{r^2} $. In  \eqref{deltawdw} the integration in $ r $ can be performed and we can re-express the result {in the form of an integration} on the null boundary using the trick  presented in details in section \ref{secdwdw}. In this method the total derivative in $ r $ is projected on null boundary using Gauss's theorem. The result is as follows 
\begin{equation}\label{deltawdw1}
I_{\delta \text{WdW}}=-2\sigma\int_{\text{WdW}} {\rm d}\la{\rm d}^2\Omega\, \sqrt{q}\,\frac{n^{\alpha}\nabla_{\alpha }\delta\Psi}{\sqrt{f(r)}}\ \left(\frac{r}{L^2} -\frac{Q^2}{r^3}\right),
\end{equation}
where $ n^{\alpha} $ is the normal to $ r=\text{const}$ surface. Using $ n^{\alpha}\partial_\alpha=\sqrt{f(r)}\partial_{r} $ and \eqref{deltapsi} we find: 
\begin{equation}\label{deltawdw2}
I_{\delta \text{WdW}}=-\sigma\int_{\text{WdW}} {\rm d}\la{\rm d}^2\Omega\, \sqrt{q}\,\deltabar\mu_{1} \left(\frac{r}{L^2} - \frac{Q^2}{r^3}\right).
\end{equation}
 Adding up  the  results from \cref{bulk1,deltawdw2,bultmatter,jointt1,dIk,dCT}  and using the expression \eqref{fsol} for $f(r)${,} we will get the result \eqref{firstlawblackhole} easily. Note that when integrating last term in \cref{bulk1}, a further sign is created which when combined with $ \sigma $, leads to $ \varepsilon $.

\bibliographystyle{hunsrt}
\bibliography{Refrences1}
\end{document}